\documentclass{article}

\usepackage{microtype}
\usepackage{graphicx}
\usepackage{subfigure}
\usepackage{booktabs} 

\usepackage{hyperref}
\usepackage{array}

\usepackage[accepted]{icml2023}

\usepackage{amsmath}
\usepackage{amssymb}
\usepackage{mathtools}
\usepackage{amsthm}

\usepackage[capitalize,noabbrev]{cleveref}

\theoremstyle{plain}
\newtheorem{theorem}{Theorem}[section]
\newtheorem{proposition}[theorem]{Proposition}

\theoremstyle{definition}
\newtheorem{definition}[theorem]{Definition}

\theoremstyle{remark}

\newcommand{\ourmethod}{\texttt{GOAL}}
\newcommand{\ourconvmethod}{\texttt{CGC}}

\newcommand{\hlt}{\textcolor{black}}
\usepackage{enumitem}

\usepackage[textsize=tiny]{todonotes}

\icmltitlerunning{Graph Complementary Learning}

\begin{document}

\twocolumn[
\icmltitle{Finding the Missing-half: Graph Complementary Learning for Homophily-prone and Heterophily-prone Graphs}




\begin{icmlauthorlist}
\icmlauthor{Yizhen Zheng}{monash}
\icmlauthor{He Zhang}{monash}
\icmlauthor{Vincent CS Lee}{monash}
\icmlauthor{Yu Zheng}{la_trobe}
\icmlauthor{Xiao Wang}{beihang}
\icmlauthor{Shirui Pan}{griffith}

\end{icmlauthorlist}

\icmlaffiliation{monash}{Monash University}
\icmlaffiliation{beihang}{Beihang University}
\icmlaffiliation{la_trobe}{La Trobe University}
\icmlaffiliation{griffith}{Griffith University}

\icmlcorrespondingauthor{Shirui Pan}{s.pan@griffith.edu.au}
\icmlcorrespondingauthor{Yizhen Zheng}{yizhen.zheng1@monash.edu}

\icmlkeywords{Machine Learning, ICML}

\vskip 0.3in
]



\printAffiliationsAndNotice{} 

\begin{abstract}
\label{abstract}
 Real-world graphs generally have only one kind of tendency in their connections. These connections are either homophily-prone or heterophily-prone. While graphs with homophily-prone edges tend to connect nodes with the same class (i.e., intra-class nodes), heterophily-prone edges tend to build relationships between nodes with different classes (i.e., inter-class nodes).
 Existing GNNs only take the original graph 
 during training. The problem with this approach is that it forgets to take into consideration the ``missing-half" structural information, that is, heterophily-prone topology for homophily-prone graphs and homophily-prone topology for heterophily-prone graphs. 
 In our paper, we introduce \textit{\textbf{G}raph c\textbf{O}mplement\textbf{A}ry \textbf{L}earning}, namely \ourmethod, which consists of two components: graph complementation and complemented graph convolution. The first component finds the missing-half structural information for a given graph to complement it. The complemented graph has two sets of graphs including both homophily- and heterophily-prone topology. In the latter component, to handle complemented graphs, we design a new graph convolution from the perspective of optimisation. The experiment results show that~\ourmethod\ consistently outperforms all baselines in eight real-world datasets.
 
\end{abstract}

\section{Introduction}
\label{sec:intro}

\begin{figure}[t]
    \centering
    \includegraphics[scale = 0.48]{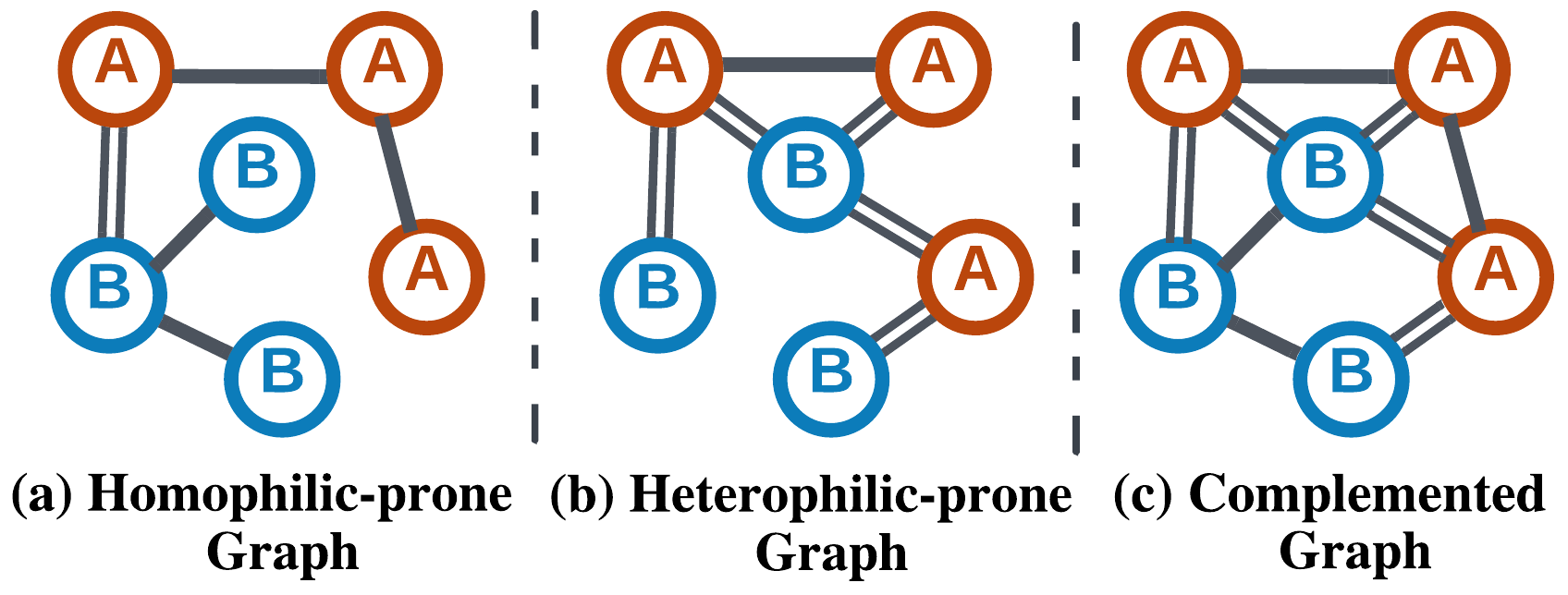}
    \vspace{-5mm}
    \caption{Example illustration of a homophily-prone graph, a heterophily-prone graph and a complemented graph. The single lines represent intra-class edges, while the doubled lines represent inter-class edges. There are two classes of nodes in these graphs, which are A class (red nodes) and B class (blue nodes).}
    \label{fig:homo_heter_all}

\end{figure}

Graph Neural Networks (GNNs) have gained much attention in recent years due to their promising performance on various downstream tasks, such as node clustering~\cite{liuyue_DCRN, liuyue_survey}, link prediction~\cite{luo2023graph}, and node classification~\cite{DBLP:conf/nips/XiongZPP0S22, liu2023beyond, DBLP:conf/ijcai/ZhuPL0YL22, zheng2022rethinking, zheng2022unifying}. In addition, they can be applied to many real-world applications, e.g., anomaly detection~\cite{ liu2023good, zheng2021heterogeneous}, neural architecture search~\cite{zheng2023auto, zheng2022multi}, knowledge graphs~\cite{luo2023npfkgc}, time series forecasting~\cite{jin2022multivariate, jin2022neural} and system security~\cite{ZhangYZP22, ZhangWY0WYP21, abs-2301-12951, abs-2205-07424,tan2023federated}. Despite the success of existing GNNs, they overlook the ``missing-half'' structural information of the given graph, that is, heterophily-prone edges for homophily-prone graphs and homophily-prone edges for heterophily-prone graphs. This is because they only use the given real-world graph during training, which normally has only one type of connection: homophily-prone or heterophily-prone. Figure~\ref{fig:homo_heter_all}(a) and (b) show examples of a homophily-prone graph and a  heterophily-prone graph, respectively. In particular, edges in a homophily-prone graph tend to connect intra-class nodes, whereas edges in a heterophily-prone graph tend to connect inter-class nodes. 


\begin{figure*}[t]
    \centering
\includegraphics[scale = 0.383]{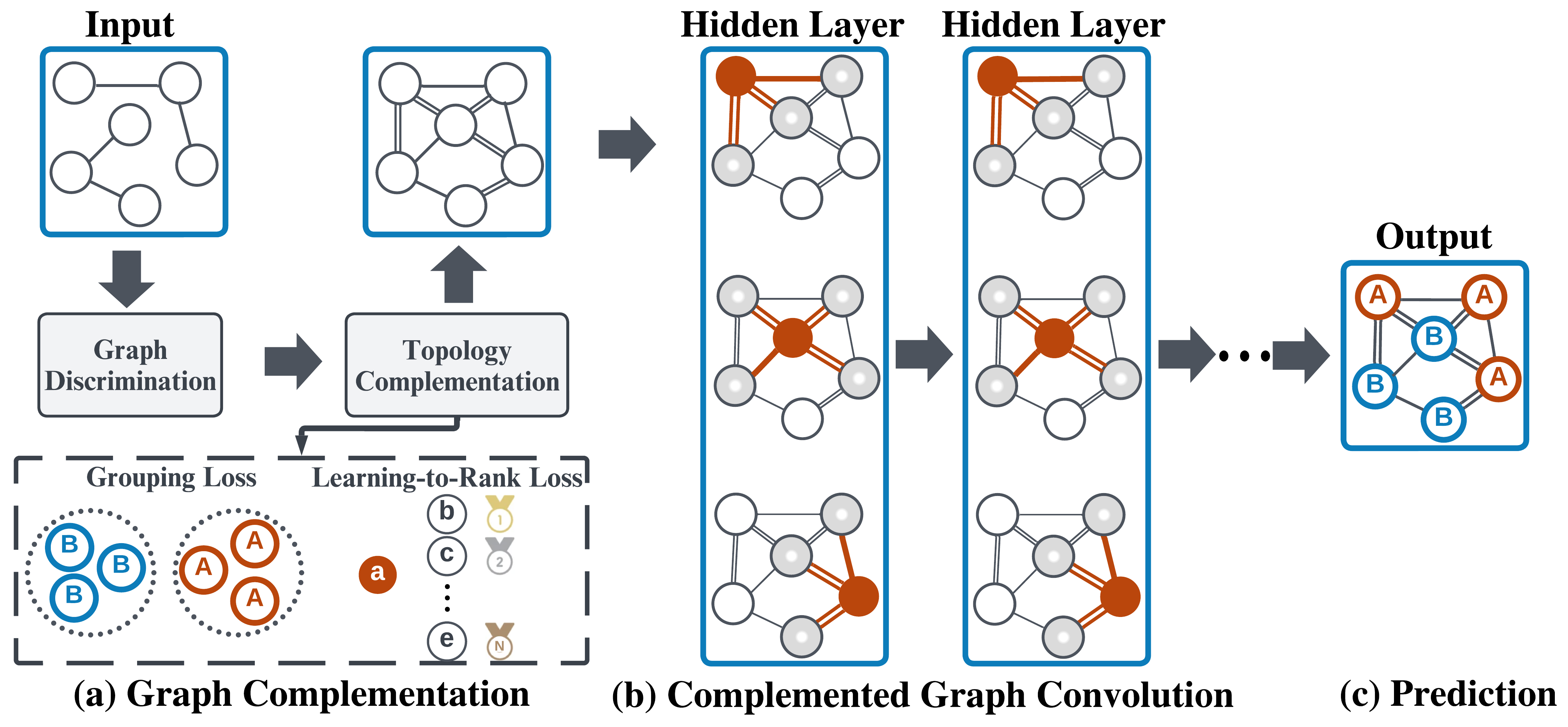}
    \caption{The overall architecture of \ourmethod. In graph complementation (a), we first perform graph discrimination to determine which part of structural information is missing from the input graph. Then, we conduct topology complementation to find and complement the input graph with another set of graph topology. This process utilises two losses including grouping loss and learning-to-rank loss. While the first loss aims to cluster intra-class nodes, the latter one attempts to predict the correct ranking of nodes in a ranking list for a target node (e.g., node $a$ in the figure). In complemented graph convolution, we conduct a newly-designed graph convolution to utilise both heterophily- and homophily-prone connections. During convolution, each node interacts with both its homophily- and heterophily-connected neighbours. The edges involved in the convolution are red-coloured and bold. Here, the single and the doubled lines indicate two different types of connections. Grey neighbouring nodes with a white centre point are involved in the convolution process of the red-coloured target node. Finally, we obtain the output prediction for each node, e.g., A or B class.}
    \label{fig:overall}

\end{figure*}


Adding the missing-half topology to a graph can enrich its structural information abundantly. With this information, we can intuitively know for each node both its intra-class and inter-class counterparts. We can then assign each node to be closer to their intra-class peers during training, and the reverse for their inter-class peers. Thus, a natural question is ``Can we find the missing-half structural information to complement real-world graphs and utilise them?''

To overcome this problem, we introduce \textit{\textbf{G}raph c\textbf{O}mplement\textbf{A}ry \textbf{L}earning}, namely \ourmethod, which can complement the given graph with the ``missing-half" topology and utilise the complemented graph for model training. Figure~\ref{fig:homo_heter_all}(c) presents an example of the complemented graph that have two set of edges including both homophily- and heterophily-prone linkages. In order to conduct Graph Complementary Learning, there are two main challenges: \textbf{(1)} \textit{How to design a model to complement the given graph?} and \textbf{(2)} \textit{How to design a new graph convolution to handle complemented graphs with both homophily- and heterophily-prone topology?}

We address the first challenge by proposing a two-step pipeline, which includes graph discrimination and topology complementation. Graph discrimination uses Kolmogorov-Smirnov statistics \cite{lilliefors1967kolmogorov} to determine which kind of connection is missing from a graph. Then, we complement a graph using a model incorporating both grouping and learning-to-rank losses. For the second challenge, we design a proper graph convolution for complemented graphs from the perspective of optimisation. The newly-designed graph convolution can extract valuable information from both homophily- and heterophily-prone topology.

The overall architecture of Graph Complementary Learning is presented in Figure~\ref{fig:overall}. It consists of two main components: graph complementation and complemented graph convolution. While the first component finds the missing-half topology for a graph, the latter one handles the complemented graph having two sets of topology. These two components are detailed in Sections~\ref{sec:method_gcl} and ~\ref{sec:method_agc}.

The contributions of our work are threefold:
\begin{itemize}
    \item To the best of our knowledge, we are the first to introduce Graph Complementary Learning, which finds the ``missing-half'' topology to complement graphs and utilise them.
    \item We propose a graph convolution strategy that can handle graphs with both homophily- and heterophily-prone topology. 
    \item Our proposed \ourmethod\ outperforms baselines on eight real-world datasets, which validates its superiority.
\end{itemize}

\section{Related Work}
\label{sec:re_work}
Graph neural networks (GNNs) can be categorised into spectral-based GNNs and spatial-based GNNs. In the spectral domain, spectral GNNs design spectral graph filters based on spectral graph theory. \citet{bruna2013spectral} first defines graph convolution filters as functions of eigenvalues of the graph Laplacian matrix. 
In ChebyNet~\cite{defferrard2016convolutional}, a spectral filter function is approximated by the Chebyshev polynomial. GPR-GNN~\cite{chien2020adaptive} approximates graph convolutions using the Monomial basis, which can derive either low or high-pass filters. Bridging spectral- and spatial-based GNNs, GCN~\cite{kipf2016semi} simplifies ChebyNet~\cite{defferrard2016convolutional} with the first-order approximation and can be interpreted from the spatial perspective. 


In addition, spatial-based methods have gained popularity in recent years due to their efficiency, generality, and flexibility~\cite{wu2020comprehensive}. Their basic idea is that graph signals can be propagated with graph structures iteratively, which broadens a node neighbourhood. For example, GCN~\cite{kipf2016semi} can be interpreted as aggregating nearest-neighbouring graph signals to generate updated node embeddings. SGC~\cite{wu2019simplifying} simplifies GCN by decoupling message passing and feature propagation. In addition, it removes the non-linearity in GCN. GAT~\cite{velivckovic2017graph} assigns learnable weights in the message-passing process.
APPNP~\cite{klicpera2019diffusion} and GDC~\cite{klicpera2019diffusion} further enrich the signal propagation with personalised PageRank and graph diffusion. 


Results from the above approaches have been promising. 
However, they directly use the original graph as input, neglecting the ``missing-half" topology. To complement the missing type of connection for the original graph, we introduce Graph Complementary Learning, namely ~\ourmethod, which can complement graphs with ``missing-half'' topology and process these complemented graphs.

\begin{figure*}[t]
    \centering
    \includegraphics[scale = 0.275]{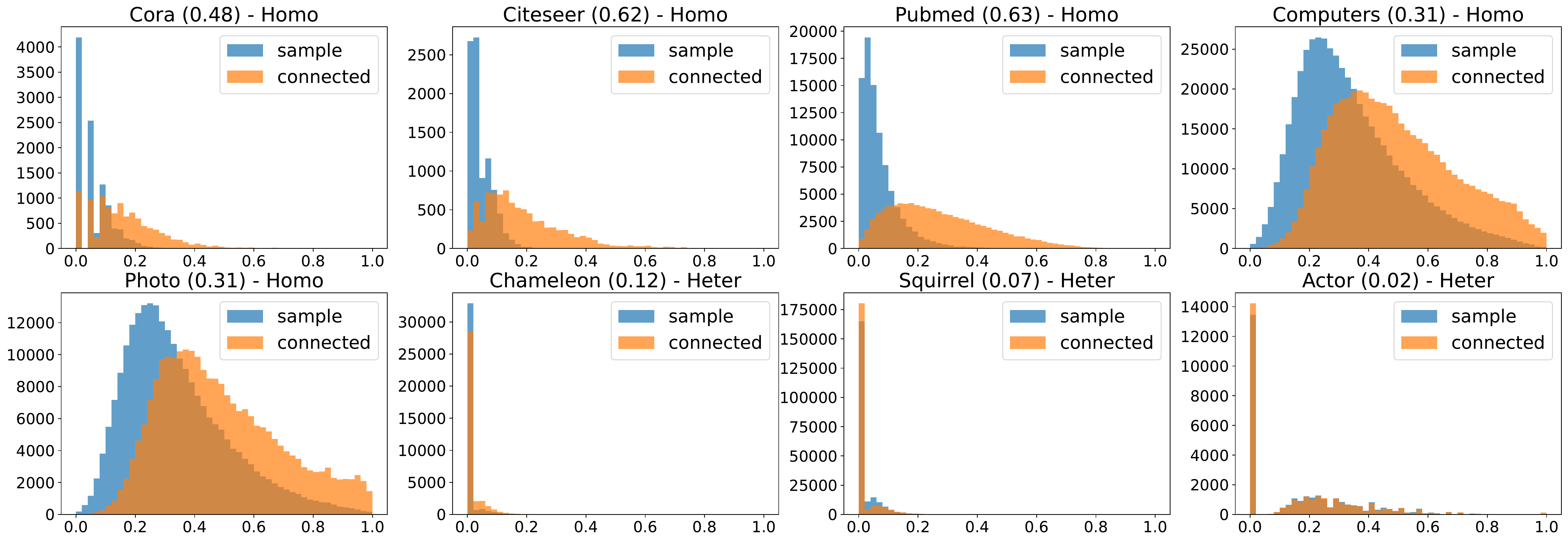}
    \caption{Here is a comparison of the CSDs between connected and randomly sampled node pairs. The number in the bracket after each dataset is the K-S statistics of the two distributions. ``Homo'' and ``Heter'' represent homophily- and heterophily-prone graphs, respectively.}
    \label{fig:cosine comp}
    \vspace{-2mm}
\end{figure*}

\section{Preliminaries}
\label{sec:pre}
\textbf{Notations.} Let $\normalfont{\mathcal{G} = \{\textbf{X} \in \mathbb{R}^{N\times D}, \textbf{A} \in \mathbb{R}^{N\times N}\}}$ denote an undirected and attributed graph, where $\textbf{A}$ is an adjacency matrix with $N$ nodes and $\textbf{X}$ is the feature matrix with dimension $D$. $\textbf{A}$ encapsulates the structural information of $\mathcal{G}$, in which a non-zero entry indicates there is a connection between two nodes (e.g., node $i$ and $j$ if $\textbf{A}_{i,j} = 1$). $(i,j) \in \mathcal{E}$ denotes a edge connecting node $i$ and node $j$ in the set of all edges $\mathcal{E}$. The number of edges in $\mathcal{E}$ is $E$. A diagonal degree matrix is defined as $\textbf{D}$, whose diagonal has values of $\textbf{D}_{i,i} = \sum_j \textbf{A}_{i,j}$. 
The normalised adjacency matrix is $\hat{\textbf{A}} = \textbf{D}^{-\frac{1}{2}}\textbf{A}\textbf{D}^{-\frac{1}{2}}$. The normalised Laplacian matrix $\hat{\textbf{L}} = \textbf{I} - \hat{\textbf{A}}$ is a symmetric and positive semi-definite matrix whose eigenvalues $\lambda$ are in range $[0, 2]$ ~\cite{chien2020adaptive}.


\textbf{Problem Formulation.}
The focus of this study is node classification tasks, which predict the class label for each node. We aim to learn a mapping $f_{\theta}(\textbf{X}, \textbf{A}) \rightarrow \textbf{Y} \in \mathbb{R}^{N \times C}$, where $f_{\theta}(\cdot)$ is a learnable function to generate a prediction matrix $\textbf{Y}$ and $C$ is the number of classes. The class with the highest value in each row of $\textbf{Y}$ will be considered as the predicted class nodes.

\vspace{2mm}

\section{Graph Complementation}
\label{sec:method_gcl}
As shown in Figure \ref{fig:overall} (a), this section presents the graph complementation module to answer ``\textit{How to design a model to complement the given graph?}" This module includes graph discrimination and topology complementation.



\subsection{Graph Discrimination}
\label{sec:method_gcl:gd}
Graph discrimination is used to determine what kind of structural information is missing from a graph. As an example, the homophily-prone Cora graph lacks the set of heterophily-prone edges. Thus, it needs to be complemented with this type of connection.

\textbf{Experiment Setup.} To distinguish homophily- and heterophily-prone datasets, we conducted an experiment to compare the cosine similarity distribution (CSD) between connected node pairs and randomly sampled node pairs. 
The cosine similarity between the nodes in a node pair $(v_i, v_j)$ can be calculated as follows:
 \begin{equation}
     cos(v_i, v_j) = \frac{\textbf{x}_i \cdot \textbf{x}_j}{\parallel \textbf{x}_i \parallel \parallel \textbf{x}_j \parallel},
 \end{equation}
 where $cos(\cdot)$ takes the features of two nodes as input and outputs the cosine similarity between the two nodes. $\textbf{x}_i \in \mathbb{R}^{1 \times D}$ is the raw feature of node $i$, and $\parallel \textbf{x}_i \parallel$ means the norm of $\textbf{x}_i$.

The cosine similarity distribution (CSD) of connected node pairs is denoted as $\textbf{d}_1$ (i.e., $\{cos(v_i, v_j)|(v_i, v_j) \in \mathcal{E}_{\mathcal{G}}$\}), where $\mathcal{E}_{\mathcal{G}}$ is the set of edges in graph $\mathcal{G}$. The CSD of randomly sampled node pairs $\textbf{d}_2 = \{cos(v_i, v_j)|(v_i, v_j) \in \mathcal{E}_{\mathcal{S}}\}$, where $\mathcal{E}_{\mathcal{S}}$ are uniformly sampled between any two nodes and its size is the same as $\textbf{d}_1$, i.e., $|\mathcal{E}_{\mathcal{G}}| = |\mathcal{E}_{\mathcal{S}}|$. 

\textbf{Observations.} As shown in Figure~\ref{fig:cosine comp}, we compare the CSDs between connected and randomly sampled node pairs on eight datasets. Our key observations are as follows.

(1) \textit{ There is an obvious CSD difference between the connected and random node pairs in homophily-prone graphs}.
Based on the visualisations in Figure~\ref{fig:cosine comp} (e.g., Cora dataset), we observe that the CSDs of connected node pairs in homophily-prone graphs differ greatly from those of randomly sampled node pairs. Specifically, CSDs of connected nodes are normally right-shifted and flatter than those of sampled node pairs in homophily-prone graphs. The right-shiftiness is reasonable because homophily-prone linkages tend to connect intra-class nodes that have a higher similarity. In addition, homophily-prone graphs exclude many inter-class node pairs with low similarity, alleviating the positive skewness (i.e., right long tail) and causing the distribution to be flatter. 

(2) \textit{heterophily-prone graphs have similar CSDs on connected and random node pairs.}
As shown in Figure~\ref{fig:cosine comp} (e.g., Chameleon dataset), the CSDs of the connected node pairs in heterophily-prone graphs have similar distributions to randomly sampled node pairs. The reason for this, we suggest, is that randomly sampled pairs of nodes tend to have mainly heterophilic relationships. As shown in Figure~\ref{fig:cosine comp}, the CSDs of randomly sampled pairs across all datasets are heavily positive-skewed, which means most sampled pairs have very low cosine similarity and are likely heterophilic.




\paragraph{Method.}
Based on the above observations, we propose a method to distinguish homophily- and heterophily-prone datasets based on the CSD difference between connected node pairs and randomly sampled node pairs. Specifically, we use Kolmogorov-Smirnov statistics (K-S statistics)~\cite{lilliefors1967kolmogorov} to evaluate the difference in distribution between connected and random node pairs. A desirable property of K-S statistics is that it ranges from 0 to 1, which allows fair comparison between different datasets. Based on this measure, we set a threshold to determine which kind of connection is likely missing. K-S statistics can be calculated with the following formula:
\begin{equation}
    KS(\textbf{d}_1, \textbf{d}_2) = max \parallel F(\textbf{d}_1) - F(\textbf{d}_2) \parallel,
\end{equation}
where $KS(\cdot)$ takes two distributions as input and output K-S statistics. $F(\cdot)$ is a function to calculate the cumulative distribution function (CDF) of the given distribution. $F(\textbf{d}_1)$ is the sample CDF and $F(\textbf{d}_2)$ is the reference CDF. In our case, $\textbf{d}_1$ and $\textbf{d}_2$ are the CSDs of connected nodes and randomly sampled nodes, respectively.

We choose 0.2 as the threshold for K-S statistics. If the K-S statistics of a graph exceed 0.2, it will be considered homophily-prone; otherwise, it will be heterophily-prone. We choose 0.2 as it indicates a clear/distinguishable difference between two distributions as suggested in \cite{da2021survey}. For each dataset, we randomly sampled ten sets of random node pairs and calculate the K-S statistics between the CSDs of these sets and connected nodes. The average results are shown in Figure~\ref{fig:cosine comp} (in brackets after each dataset name). From the figure, we conclude that this approach successfully distinguishes all graphs (i.e., Cora, Citeseer, Pubmed, Computers and Photo are homophily-prone / Chameleon, Squirrel, and Actor are heterophily-prone) in terms of their tendency in connections.

\subsection{Topology Complementation}
In topology complementation, we aim to find the missing-half structural information and complement the given graph with additional structural information. To achieve this, we train a model with two losses: grouping and learning-to-rank loss. The overall loss can be formulated as follows:
\begin{equation}
    \mathcal{L} = \mathcal{L}_{grp} + \mathcal{L}_{rank}, 
    \label{eq:overall loss}
\end{equation}
where $\mathcal{L}_{grp}$ is the grouping loss and $\mathcal{L}_{rank}$ is the learning-to-rank loss.

\paragraph{Grouping Loss.} Grouping loss can help a model group intra-class nodes together while separating inter-class nodes. To do this, we first obtain node embeddings with a backbone GNN encoder, e.g., GAT~\cite{velivckovic2017graph}. Then, node embeddings are input to a (Multi-layer Perceptron) MLP model. Afterwards, the MLP model needs to increase the similarity between intra-class node pairs, while doing the opposite for inter-class node pairs by optimising the grouping loss. Specifically, we first obtain all intra-class node pairs as the positive set $\mathbb{P} = \{(v_i, v_j) | Y_{v_i} = Y_{v_j}\}$, while all inter-class node pairs as the negative set $\mathbb{N} = \{(v_i, v_j) | Y_{v_i} \neq Y_{v_j}\}$. Here, $v_i$ and $Y_{v_i}$ represent node $i$ and its label, respectively. The whole process can be formulated as follows:
\begin{equation}
        \rm \textbf{Z}_{gc} = MLP(\textbf{Z}_{gnn});\ \textbf{Z}_{gnn}  = GNN(\mathcal{G}),
\end{equation}
\begin{equation}
\begin{aligned}
    \mathcal{L}_{pos} &= - log (sig(\frac{\sum_{(i,j) \in \mathbb{P}}\textbf{Z}_{gc}^T[v_i] \cdot \textbf{Z}_{gc}[v_j]}{|\mathbb{P}|}) + \epsilon),\\
     \mathcal{L}_{neg} &= -log (1 - sig(\frac{\sum_{(i,j)\in \mathbb{N}}\textbf{Z}_{gc}^T[v_i] \cdot \textbf{Z}_{gc}[v_j]}{|\mathbb{N}|} + \epsilon)),
\end{aligned}
\end{equation}
\begin{equation}
         \mathcal{L}_{grp} = \mathcal{L}_{pos} + \mathcal{L}_{neg},
         \label{eq:group loss}
 \end{equation}
 where $\cdot$ represents matrix multiplication, $\textbf{Z}_{gnn}$ is the output embedding of the GNN encoder taking $\mathcal{G}$ as input, $\textbf{Z}_{gc}$ is the output embedding of the MLP model, $sig(\cdot)$ is the sigmoid function, $\epsilon$ is a tiny number to prevent computation error. Note that $\textbf{Z}_{gnn}$ is obtained by the \hlt{pretrained GNN encoder with cross-entropy loss}. We take the embedding before the prediction layer as $\textbf{Z}_{gnn}$.

\begin{figure}
    \centering
    \includegraphics[scale = 0.55]{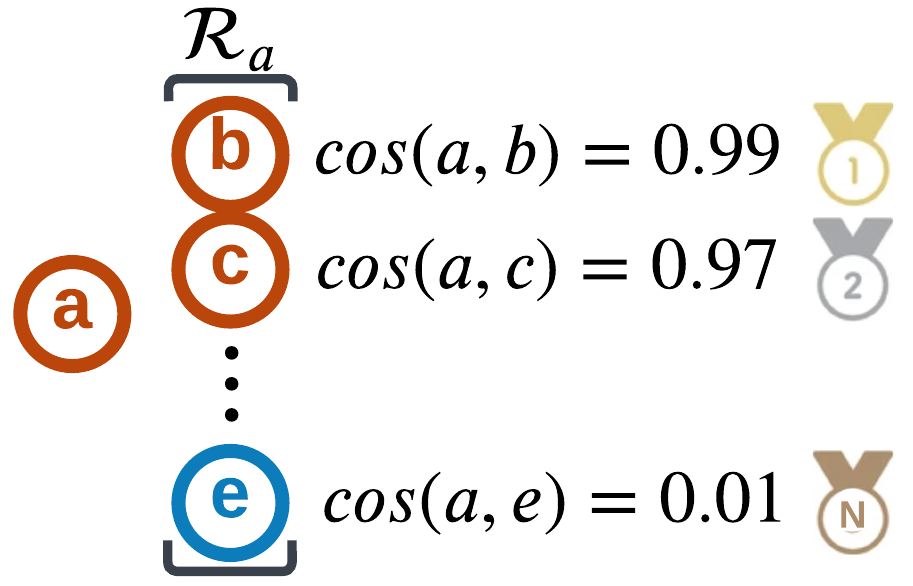}
    \caption{An example of ranking list $\mathcal{R}_a$ for node $a$. Red and blue coloured nodes indicate two different classes of nodes.}
    \label{fig:ranking}
\end{figure}

\paragraph{Learning-to-Rank Loss.}
In this paper, Learning-to-Rank aims to predict the ranking of other nodes based on a given node and their cosine similarity. 
Concretely, for a target node, \hlt{we consider intra-class nodes with higher cosine similarity having higher ranks}, while inter-class with lower cosine similarity have lower ranks. Each node will have its corresponding ranking list and an example is illustrated in Figure~\ref{fig:ranking}. With this loss, the model is expected to predict the correct ranking of nodes in the ranking list for each target node.

We adopt ListNet loss ~\cite{cao2007learning}, a listwise learning-to-rank approach, to conduct the learning-to-rank task. It can optimise the ranking of a list of items by minimising the cross entropy between the predicted and true ranking of items in a ranking list. To build the ranking list $\mathcal{R}_i$ for a target node $i$, we first sort its intra-class nodes in the training set based on cosine similarity and obtain $L^{intra}_{i} = \{v^{1}_j, ... v^{Q}_j| Y_{v^{q}_j} = Y_{v_i}; cos(v_i, v^{q}_j) > cos(v_i, v^{q + 1}_j)\}$. Here, $1 \leq q	\leq Q$ and $Q$ is the number of intra-class nodes for node $i$. $cos(\cdot)$ is the cosine similarity function. Similarly, we can obtain $L^{inter}_{i} = \{v^{1}_j, ... v^{E}_j| Y_{v^{e}_j} \neq Y_{v_i} ; cos(v_i, v^{e}_j) > cos(v_i, v^{e + 1}_j)\}$, where \hlt{$1 \leq e	\leq E$} and $E$ is the number of inter-class nodes. Finally, we can obtain the ranking list $\mathcal{R}_i = \{L^{intra}_{i}[1:K] \parallel L^{inter}_{i}[-K:-1]\}$, which includes top $K$ most similar intra-class and least similar inter-class nodes in the training set for node $i$. $\parallel$ means concatenation. The position of each node in $\mathcal{R}_i$ determines its true ranking. The ListNet loss can be formulated as follows:
\begin{equation}
\begin{aligned}
    \hat{\textbf{y}}_{v_i} &= \textbf{Z}_{gc}[i]^T \cdot \textbf{Z}_{gc}[\mathcal{R}_i], \\
    \mathcal{L}_{rank} &= -\sum^B_{i = 1}(p(\textbf{y}_{v_i})log(p(\hat{\textbf{y}}_{v_i}))),
\end{aligned}
    \label{eq:rank loss}
\end{equation}
where $B$ represents the size of the training set, $p(\cdot)$ is the softmax function, $\hat{\textbf{y}}_{v_i}$ indicates the predicted score list for the nodes in ranking list $\mathcal{R}_i$, and $\textbf{y}_{v_i}$ denotes the \hlt{ground truth} obtained from $\mathcal{R}_i$~\cite{cao2007learning}.

\paragraph{Topology Complementation.} 
\label{par:graph compl} 
After the topology complementation module is well trained by minimising $\mathcal{L} = \mathcal{L}_{grp} + \mathcal{L}_{rank}$, it can generate an additional topology for a graph based on its graph discrimination result in Section \ref{sec:method_gcl:gd}.
For graphs discriminated as homophily-prone graphs, additional heterophily-prone edges are built by finding the $K_{heter}$ least similar nodes for each node. Specifically, we compute the dot product similarity between all node pairs via $\textbf{Z}_{gc} \cdot \textbf{Z}_{gc}^T$. Then, we find those top $K_{heter}$ least similar nodes of each node and connect them to it. In contrast, for graphs discriminated as heterophily-prone graphs, we find the $K_{homo}$ most similar nodes for each node to build the homophily-prone topology.

\section{Complemented Graph Convolution}
\label{sec:method_agc}

Given the complemented graphs, this section aims to answer ``\textit{how to design a new graph convolution to handle complemented graphs with both homophily- and heterophily-prone topology?}" 
In this section, we first present the convolution designed for complemented graphs, followed by how to derive it from our optimisation goal, which is proposed for simultaneously digging homophily- and heterophily-prone topology information.

\subsection{Complemented Graph Convolution Solution}

In this paper, we propose \textbf{C}omplemented \textbf{G}raph \textbf{C}onvolution (\ourconvmethod) from the perspective of optimisation.
Specifically, the \ourconvmethod\ operation can be formulated as follows:
\begin{equation}
    \textbf{H}^{l + 1} = \sigma((\alpha \textbf{I} + \beta \hat{\textbf{A}}_{o}- \gamma \hat{\textbf{A}}_{t} - \delta\hat{\textbf{A}}_{to})\textbf{H}^l\textbf{W}^l),
    \label{eq:agc conv}
\end{equation}
where $\textbf{H}^{l} \in \mathbb{R}^{N\times D'}$ is node representations at $l$-th layer ($\textbf{H}^{(0)} = \textbf{X}$), $\textbf{W}^l \in \mathbb{R}^{D'\times D'}$ is a weight matrix at $l$-th layer $(\textbf{W}^0 \in \mathbb{R}^{D \times D'})$, $\hat{\textbf{A}}_{o}$ represents the normalised homophily-half of graph, $\hat{\textbf{A}}_{t}$ represents the normalised heterophily-half of graph, and $\hat{\textbf{A}}_{to} = \hat{\textbf{A}}_{t}\hat{\textbf{A}}_{o}$ is the combined topology. $\alpha$, $\beta$, $\gamma$, and $\delta$ are four parameters that determine the strength of self, homophily-prone, heterophily-prone and combined information, respectively. The combined information involves both homophily- and heterophily-prone topology. These parameters are ranged from $[0,5]$. 

\subsection{\ourconvmethod : Interpretation and Derivation from Optimisation Perspectives}

\label{sec:optimisation agc}
To take full advantage of both  homophily- and heterophily-prone structural information, we propose to \textit{maximise the similarity between connected nodes from homophily-prone edges while doing the opposite with heterophily-prone edges}.
Next, we first present the optimisation objective which provides an interpretation of the underlying mechanism and then show how we derive our \ourconvmethod\ from it. An interpretation of our \ourconvmethod\  from the graph signal processing perspective is also given in Appendix~\ref{sec:appendix spectral}.

According to a recent study on interpreting and unifying GNNs \cite{zhu2021interpreting}, existing various GNN architectures (e.g., GCN) can be derived from a framework with the optimisation perspective.
In this paper, we follow the same pipeline and propose the following optimisation objective to conduct learning on complement graphs, i.e.,
\begin{align}
\begin{aligned}
    \mathcal{O}_{\ourconvmethod} &= \underset{\textbf{H}}{min}\{
    tr(\textbf{H}^T(3\textbf{I} - \hat{\textbf{A}}_{o}+\hat{\textbf{A}}_{t}-\hat{\textbf{A}}_{t}\hat{\textbf{A}}_{o})\textbf{H})\},
    \label{eq:all obj}
\end{aligned}
\end{align}
where $tr(\cdot)$ indicates the trace of a matrix. 
As shown in the following, this objective can be regarded as a combination of three objectives $\mathcal{O}_{o}, \mathcal{O}_{t}$, and $\mathcal{O}_{c}$, which respectively utilise homophily- ($\hat{\textbf{A}}_o$), heterophily-prone ($\hat{\textbf{A}}_t$) and combined topology ($\hat{\textbf{A}}_{to}$) to implement our intuition.
Concretely,
\begin{equation}
\label{eq:homo obj}
\begin{aligned}
    \mathcal{O}_{\ourconvmethod} &= \mathcal{O}_{o} + \mathcal{O}_{t} + \mathcal{O}_{c},\\
    \mathcal{O}_{o} &= \underset{\textbf{H}}{min}\{tr(\textbf{H}^T(\textbf{I} - \hat{\textbf{A}}_{o})\textbf{H})\} \\ & = \frac{1}{2} \sum^E_{(i,j) \in \mathcal{E}} \hat{\textbf{A}}_{o}[i,j] \parallel\textbf{H}_i-\textbf{H}_j\parallel^2,\\
    \mathcal{O}_{t} &= \underset{\textbf{H}}{min}\{tr(\textbf{H}^T(\textbf{I} 
 + \hat{\textbf{A}}_{t})\textbf{H})\} \\ &= \frac{1}{2} \sum^E_{(i,j) \in \mathcal{E}} \hat{\textbf{A}}_t[i,j] \parallel\textbf{H}_i + \textbf{H}_j\parallel^2,\\
   \mathcal{O}_{c} &= \underset{\textbf{H}}{min}\{ tr(\textbf{H}^T(\textbf{I} + \hat{\textbf{A}}_{to})\textbf{H})\}; \hat{\textbf{A}}_{to} = \hat{\textbf{A}}_{t}\hat{\textbf{A}}_{o} \\ & = \frac{1}{2} \sum^E_{(i,j) \in \mathcal{E}} \hat{\textbf{A}}_{to}[i,j] \parallel\textbf{H}_i + \tilde{\textbf{H}}_j\parallel^2.  
\end{aligned}
\end{equation}
\begin{figure}
    \centering
    \includegraphics[scale = 0.45]{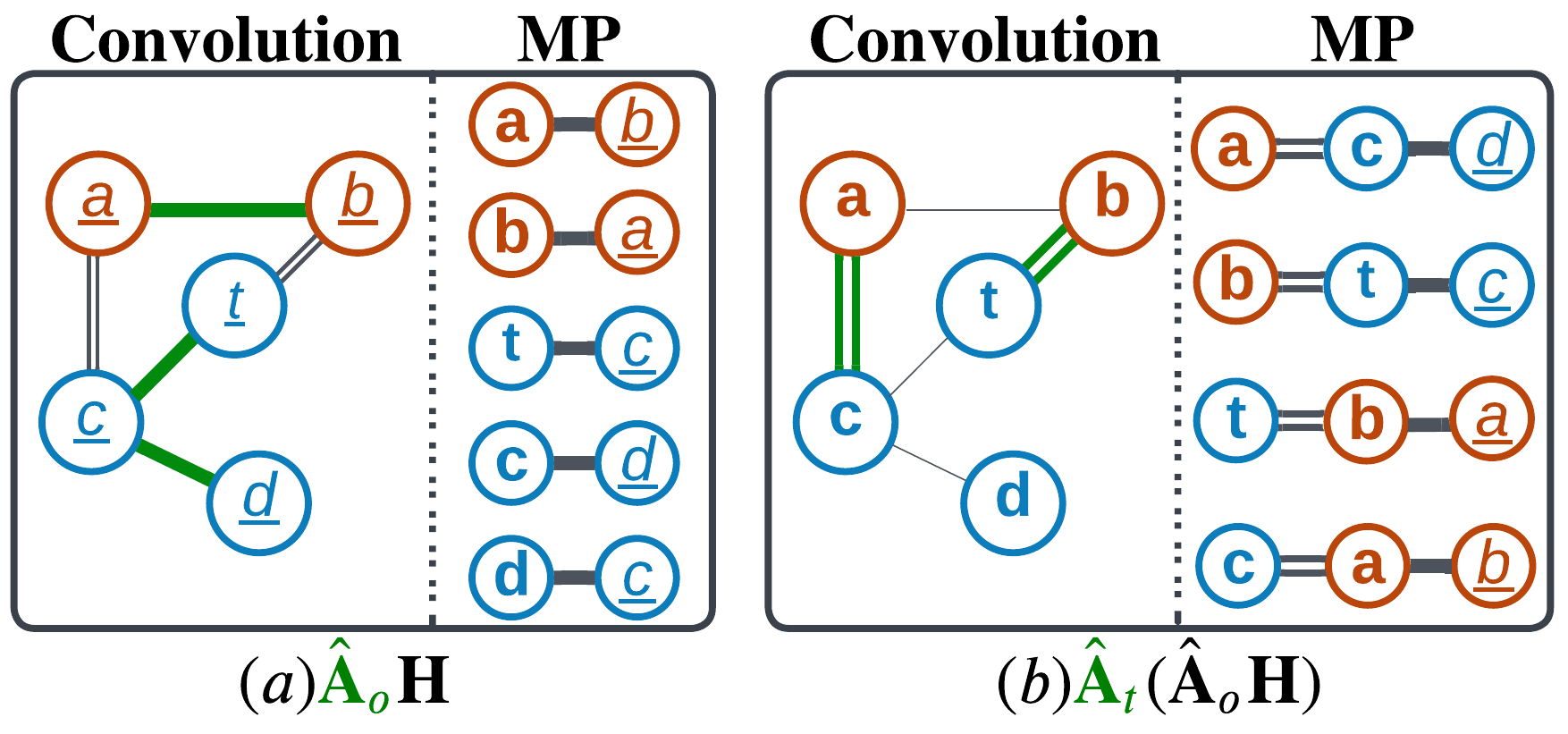}
    \caption{Illustration of $\hat{\textbf{A}}_{to}\textbf{H}$. ``MP'' means message passing chains. The left-hand side of each subfigure presents the convolution process, while the right-hand side shows message-passing chains. Edges involved in the convolution are coloured in green and bold. The underlined italic letter represents the initial node embedding. After convolution in (a), the convoluted nodes are represented with bold roman letters. The straight lines are homophily-prone connections, while the doubled lines are the opposite.}
    \label{fig:a_ot}
\end{figure}
In our optimisation objective $\mathcal{O}_{\ourconvmethod}$, $ \mathcal{O}_{o}$ aims to maximise the similarity between connected nodes from the homophilic view, while $\mathcal{O}_{t}$ and $\mathcal{O}_{c}$ are devised to minimise that from the heterophilic perspective.
Next, we will show how to convert these objectives to convolution operations, respectively.
\begin{itemize}[leftmargin=10pt]
\item $\mathcal{O}_{o}$ aims to minimise the Euclidean distance (ED) between target nodes and their homophily-prone neighbours, and it can be solved with $\frac{\partial  tr(\textbf{H}^T(\textbf{I} - \hat{\textbf{A}}_{o})\textbf{H})}{\partial \textbf{H}} = 0$ as the objective is a convex function. This is because $\hat{\textbf{L}}_{o} = \textbf{I} - \hat{\textbf{A}}_{o}$ is a positive semi-definite matrix as $\hat{\textbf{L}}_{o}$ is the laplacian matrix of $\hat{\textbf{A}}_{o}$. Therefore, we can regard $\textbf{H}^T\hat{\textbf{L}}_{o}\textbf{H}$ as the energy of the system $\hat{\textbf{L}}_{o}$ \cite{strang2006linear}. As a result, we can derive:
\begin{align}
\begin{aligned}
\hat{\textbf{L}}_{o}\textbf{H} = 0 &\to  \textbf{H} -  \hat{\textbf{A}}_{o}\textbf{H} = 0  \to \textbf{H} = \hat{\textbf{A}}_{o}\textbf{H}.
\end{aligned}
\label{eq:homo conv}
\end{align}
This result is also validated in ~\cite{zhu2021interpreting}.
\item $\mathcal{O}_{t}$ is proposed to maximise the dissimilarity between nodes and their heterophily-prone neighbours.
We can also solve it via $\frac{\partial tr(\textbf{H}^T(\textbf{I} + \hat{\textbf{A}}_{t})\textbf{H})}{\partial \textbf{H}} = 0$. This is because $\tilde{\textbf{A}}_{t} = \textbf{I} + \hat{\textbf{A}}_{t}$ is still a positive semi-definitive matrix. The eigenvalue range of a normalised adjacency matrix $\hat{\textbf{A}}_{t}$ is $[-1, 1]$ ~\cite{huang2020tackling}. Thus, $\hat{\textbf{A}}_{t}\textbf{x} = \lambda_{t} \textbf{x} \to \hat{\textbf{A}}_{t}\textbf{x} + \textbf{x} = \lambda_{t} \textbf{x} + \textbf{x} \to (\textbf{I} + \hat{\textbf{A}}_t)\textbf{x} = (\lambda_{t} + 1)\textbf{x}$, where $\textbf{x}$ is a vector, and $\lambda_{t}$ is the eigenvalue of $\hat{\textbf{A}}_{t}$. Then, we know the eigenvalues of $\tilde{\textbf{A}}_{t}$ are in the range $[0, 2]$ and $\tilde{\textbf{A}}_{t}$ is a positive semi-definite matrix. As a result, $\mathcal{O}_{t}$ is still convex and we have:
\begin{align}
    \begin{aligned}
    (\textbf{I} + \hat{\textbf{A}}_{t})\textbf{H} = 0  &\to \textbf{H} = -\hat{\textbf{A}}_{t}\textbf{H}.
    \end{aligned}
    \label{eq:heter conv}
\end{align}
\item $\mathcal{O}_{c}$ can maximise the ED between target nodes and their neighbours with respect to $\hat{\textbf{A}}_{to}$, which can be regarded as the ``enriched heterophily-prone neighbours''. 
Specifically, edges in $\hat{\textbf{A}}_{to}$ connect each node with the homophily-prone neighbours of its heterophily-prone neighbours, i.e., 
$\hat{\textbf{A}}_{to}$ connects target nodes with extended heterophily-prone neighbours.
As shown in Figure~\ref{fig:a_ot}, when computing $\hat{\textbf{A}}_{to}\textbf{H} = \hat{\textbf{A}}_{t}\hat{\textbf{A}}_{o}\textbf{H}$, all nodes first receives messages from their 1-hop homophily-prone neighbours as shown in Figure~\ref{fig:a_ot}(a). Then, these nodes receive messages from their convoluted heterophily-prone neighbours as shown in Figure~\ref{fig:a_ot}(b). 
When solving $\mathcal{O}_{c}$, $\tilde{\textbf{A}}_{to} = \textbf{I} + \hat{\textbf{A}}_{to}$ is positive semi-definite since $\rho(\tilde{\textbf{A}}_{to}) = ||\tilde{\textbf{A}}_{to}||_{2}=||\hat{\textbf{A}}_{t}\hat{\textbf{A}}_{o}||_{2}\leq||\hat{\textbf{A}}_{t}||_{2}||\hat{\textbf{A}}_{o}||_{2} = \rho(\hat{\textbf{A}}_{t}) \rho(\hat{\textbf{A}}_{o})\leq 1$, where $\rho(\cdot)$ indicates the spectral radius of a matrix. We also provide empirical evidence of this statement in Appendix~\ref{sec:appendix eigen}.
Thus, we can solve $\mathcal{O}_{c}$ via $\frac{\partial tr(\textbf{H}(\textbf{I} + \hat{\textbf{A}}_{to})\textbf{H})}{\partial \textbf{H}} = 0$:
\begin{align}
    \begin{aligned}
(\textbf{I} + \hat{\textbf{A}}_{to})\textbf{H} = 0 &\to \textbf{H} =   -\hat{\textbf{A}}_{to}\textbf{H}. \\ 
    \end{aligned}
    \label{eq:comb conv}
\end{align}
\end{itemize}

When $\mathcal{O}_{\ourconvmethod}$ is optimised, the learnt node embedding converges to $\textbf{H} = (\hat{\textbf{A}}_o - \hat{\textbf{A}}_t - \hat{\textbf{A}}_{to})\textbf{H}$. 
After involving non-learning activation and affine transformation, we obtain the generalised \ourconvmethod\ operation:
\begin{equation}
    \textbf{H} = \sigma((\mathbf{I}+\hat{\textbf{A}}_o - \hat{\textbf{A}}_t - \hat{\textbf{A}}_{to})\textbf{H} \mathbf{W}).
    \label{eq:proof agc conv}
\end{equation}
Note that Equation~\ref{eq:proof agc conv} is a special case of Equation~\ref{eq:agc conv} (i.e., $\alpha, \beta, \gamma, \delta = 1$), the item $\mathbf{I}$ here is used to maintain self-information of nodes.

The above analysis demonstrates our \ourconvmethod\ can utilise both homophily- and heterophily-prone topology simultaneously, which helps GNNs take full advantage of the topology information in complemented graphs. 
In addition, bridging the aforementioned optimisation objectives with graph spectral theory, \ourconvmethod\ can be interpreted from the spectral perspective. In light of this, we propose the following proposition to interpret our \ourconvmethod.
\begin{proposition}
\label{prop:agc conv spectral}
In our \ourconvmethod\ convolution, the message-passing among nodes (i.e., $\normalfont{\textbf{H} = (\hat{\textbf{A}}_o - \hat{\textbf{A}}_t - \hat{\textbf{A}}_{to})\textbf{H}}$) under the spatial perspective is equivalent to applying a linear low-pass spectral filter $\normalfont{\textbf{F}_{low} = 1 - \lambda}$ to the Laplacian of $\normalfont{\hat{\textbf{A}}_o}$, a linear high-pass spectral filter $\normalfont{\textbf{F}_{high} = \lambda - 1}$ to the Laplacian of $\normalfont{\hat{\textbf{A}}_t}$ and the same high-pass spectral filter to
 the Laplacian of $\normalfont{\hat{\textbf{A}}_{to}}$. 
\end{proposition}

The detailed analysis and proof for Proposition~\ref{prop:agc conv spectral} is presented in Appendix~\ref{sec:appendix spectral}. As a result, the low-frequency band of signals in the Laplacian of $\hat{\textbf{A}}_{o}$ is preserved, while the high-frequency band of signals in the Laplacian of $\hat{\textbf{A}}_{t}$ and $\hat{\textbf{A}}_{to}$ remain.

\section{Experiment}
\label{sec:experiment}

\begin{table*}
\centering
\footnotesize
\setlength{\tabcolsep}{3.8pt}
\label{tab:node classification}
\caption{Node classification for eight baselines on eight datasets. 
}
\begin{tabular}{c|cccccccc} 
\toprule
                   & \textbf{Cora}          & \textbf{Citeseer}      & \textbf{Pubmed}        & \textbf{Computer}      & \textbf{Photo}         & \textbf{Chameleon}     & \textbf{Squirrel}      & \textbf{Actor}          \\ 
\midrule
\textbf{MLP}       & 72.09 $\pm$ 0.32          & 71.67 $\pm$ 0.40          & 87.47 $\pm$ 0.14          & 83.59 $\pm$ 0.89          & 90.49 $\pm$ 0.20            & 46.55 $\pm$ 0.42          & 30.67 $\pm$ 0.52          & 28.75 $\pm$ 0.88           \\
\textbf{GCN}       & 87.50 $\pm$ 1.04          & 75.11 $\pm$ 1.12          & 87.20 $\pm$ 0.52          & 83.55 $\pm$ 0.38          & 89.30 $\pm$ 0.82          & 62.72 $\pm$ 2.09          & 47.26 $\pm$ 0.34            & 29.98 $\pm$ 1.18           \\
\textbf{GAT}       & 88.25 $\pm$ 1.22           & 75.75 $\pm$ 1.23           & 85.88 $\pm$ 0.38          & 85.36 $\pm$ 0.50           & 90.81 $\pm$ 0.22            & 62.19 $\pm$ 3.78          & 51.80 $\pm$ 1.04          & 28.17 $\pm$ 1.19           \\
\textbf{APPNP}     & 88.36 $\pm$ 0.61          & 76.03 $\pm$ 1.27           & 86.21 $\pm$ 0.25          & 88.32 $\pm$ 0.36           & 94.44 $\pm$ 0.36           & 50.88 $\pm$ 1.18           & 33.58 $\pm$ 1.00           & 29.82 $\pm$ 0.82           \\
\textbf{GraphSage} & 88.01 $\pm$ 1.29          & 75.17 $\pm$ 1.35          & 87.39 $\pm$ 0.84          & 88.54 $\pm$ 0.69          & 94.23 $\pm$ 0.62          & 58.82 $\pm$ 2.29          & 41.19 $\pm$ 0.75          & 31.76 $\pm$ 0.73           \\
 \textbf{ChebyNet}  & 87.49 $\pm$ 0.90          & 75.50 $\pm$ 0.87          & 89.05 $\pm$ 0.29           & 89.77 $\pm$ 0.36           & 95.02 $\pm$ 0.41           & 59.98 $\pm$ 1.54           & 40.18 $\pm$ 0.55           & 35.85 $\pm$ 1.05           \\
\textbf{GPR-GNN}    & 88.65 $\pm$ 0.75          & 75.70 $\pm$ 0.81          & 88.53 $\pm$ 0.30          & 87.63 $\pm$ 0.48           & 94.60 $\pm$ 0.30           & 67.96 $\pm$ 2.55           & 49.52 $\pm$ 5.00           & 30.78 $\pm$ 0.61           \\
\textbf{JKNET}     & 86.99 $\pm$ 1.60          & 75.38 $\pm$ 1.30          & 88.64 $\pm$ 0.51          & 86.97 $\pm$ 0.56          & 92.68 $\pm$ 0.58          & 64.63 $\pm$ 3.08          & 44.91 $\pm$ 1.94          & 28.48 $\pm$ 1.25           \\ 
\midrule
$\textbf{\ourmethod}$           & \textbf{88.75 $\pm$ 0.87} & \textbf{77.15 $\pm$ 0.95} & \textbf{89.25 $\pm$ 0.55} & \textbf{91.33 $\pm$ 0.38} & \textbf{95.60 $\pm$ 0.44} & \textbf{71.65 $\pm$ 1.66}          & \textbf{60.53 $\pm$ 1.60} & \textbf{36.46 $\pm$ 1.02}  \\
\bottomrule
\end{tabular}
\end{table*}

\begin{table*}
\caption{The ablation study for comparing seven variants of \ourmethod\ and \ourmethod.}
\label{tab:ablation}
\centering
\footnotesize
\setlength{\tabcolsep}{3.8pt}
\begin{tabular}{l|cccccccc} 
\toprule
& \textbf{Cora}          & \textbf{Citeseer}       & \textbf{Pubmed}        & \textbf{Computer}      & \textbf{Photo}         & \textbf{Chameleon}     & \textbf{Squirrel}      & \textbf{Actor}          \\ 
\midrule
$\textbf{\ourmethod}_{\alpha = 0}$                                           & 88.47 $\pm$ 0.92 & 76.16 $\pm$ 2.70 & 87.51 $\pm$ 0.42 & 90.51 $\pm$ 0.53 & 94.05 $\pm$ 0.70 & \textbf{71.65} $\pm$ 1.66 & \textbf{60.53} $\pm$ \textbf{1.60} & 35.71 $\pm$ 0.80  \\
$\textbf{\ourmethod}_{\beta = 0}$                                           & 87.63 $\pm$ 1.41 & 76.28 $\pm$ 0.42  & 85.66 $\pm$ 0.59 & 60.38 $\pm$ 2.52 & 82.44 $\pm$ 2.83 & 67.27 $\pm$ 2.28 & 54.68 $\pm$ 0.68 & 35.20 $\pm$ 0.77  \\
$\textbf{\ourmethod}_{\gamma = 0}$                                        & 87.69 $\pm$ 1.58 & 76.92 $\pm$ 1.10  & 88.58 $\pm$ 0.74 & \textbf{91.33 $\pm$ 0.38} & 93.76 $\pm$ 0.53 & 69.46 $\pm$ 1.63 & 56.41 $\pm$ 2.31 & 35.49 $\pm$ 0.64  \\
$\textbf{\ourmethod}_{\delta = 0}$                                           & 88.12 $\pm$ 2.14 & 76.95 $\pm$ 1.01  & \textbf{89.25 $\pm$ 0.55} & 90.71 $\pm$ 1.51 & 94.67 $\pm$ 0.68 & \textbf{71.65 $\pm$ 1.66} & \textbf{60.53 $\pm$ 1.60} & 35.31 $\pm$ 0.68  \\
$\textbf{\ourmethod}_{\beta}$                                          & 87.54 $\pm$ 1.31 & 75.54 $\pm$ 0.81  & 87.45 $\pm$ 0.35 & 89.59 $\pm$ 1.42 & 93.97 $\pm$ 0.66 & 69.08 $\pm$ 1.15 & 57.06 $\pm$ 1.29 & 35.89 $\pm$ 1.24  \\
$\textbf{\ourmethod}_{\gamma}$                                           & 85.24 $\pm$ 1.53 & 74.70 $\pm$ 2.87  & 86.01 $\pm$ 0.83 & 84.25 $\pm$ 1.16 & 91.75 $\pm$ 0.66 & 66.53 $\pm$ 2.09 & 50.12 $\pm$ 4.83 & 35.37 $\pm$ 0.52  \\
$\textbf{\ourmethod}_{\delta}$                                          & 84.83 $\pm$ 1.85 & 74.85 $\pm$ 2.93  & 83.24 $\pm$ 1.23 & 78.19 $\pm$ 5.23 & 90.54 $\pm$ 0.55 & 54.34 $\pm$ 2.81 & 25.11 $\pm$ 4.06 & 35.28 $\pm$ 0.50  \\ 
\midrule
{\textbf{\ourmethod}}  & \textbf{88.75 $\pm$ 0.87} & \textbf{77.15 $\pm$ 0.95}  & \textbf{89.25 $\pm$ 0.55} & \textbf{91.33 $\pm$ 0.38} & \textbf{95.60 $\pm$ 0.44} & \textbf{71.65 $\pm$ 1.66} & \textbf{60.53 $\pm$ 1.60} & \textbf{36.46 $\pm$ 1.02}  \\
\bottomrule
\end{tabular}
\vspace{-1mm}
\end{table*}

For experiments, we first conduct node classification tasks to evaluate the effectiveness of \ourmethod\ on eight real-world datasets. Then we conduct ablation studies of \ourmethod. 
Then, we evaluate the quality of the found missing-half topology in terms of homophily ratio for eight datasets. After that, we provide an automatic parameter-tuning method to alleviate the burden of parameter tuning. Finally, we introduce how to extend \ourmethod\ to large-scale datasets. Our code is available at \href{https://github.com/zyzisastudyreallyhardguy/GOAL-Graph-Complementary-Learning}{https://github.com/zyzisastudyreallyhardguy/GOAL-Graph-Complementary-Learning}

\subsection{Node Classification on Real-World Datasets}
In this section, we compare \ourmethod\ with eight baselines on eight real-world datasets for node classification. The selected baselines are MLP, GCN~\cite{kipf2016semi}, GAT~\cite{velivckovic2017graph}, APPNP~\cite{klicpera2018predict}, GraphSage~\cite{hamilton2017inductive},
ChebyNet~\cite{defferrard2016convolutional}, GPR-GNN~\cite{chien2020adaptive} and JKNET~\cite{chen2020simple}. The adopted datasets include three citation networks (i.e., Cora, Citeseer, and Pubmed)~\cite{yang2016revisiting}, two Amazon co-purchasing networks (i.e., Computers and Photo)~\cite{shchur2018pitfalls}, two Wikipedia graphs (i.e., Chameleon and Squirrel)~\cite{rozemberczki2021multi} and the Actor co-occurrence graph~\cite{pei2020geom}. While the first five are homophily-prone graphs, the last three are heterophily-prone. The dataset statistics and the parameter settings of \ourmethod\ are presented in Appendix \ref{sec:appendix exp}. 

For dataset split, we randomly split the node set of a dataset according to the ratio 60\%/20\%/20\% for training, validation and test set respectively. We randomly generate the split 10 times and run each baseline and \ourmethod. Experiment results including the averaged accuracy and standard deviation are summarised in Table~\ref{tab:node classification}. 

As shown in Table ~\ref{tab:node classification}, our $\ourmethod$ outperforms all baselines on all datasets, which validates the effectiveness of $\ourmethod$. Notably, \ourmethod\ is nearly 9.0\% and 3.7\% better than the most competitive baseline on Squirrel and Chameleon, respectively. The improvement can be attributed to the utilisation of the missing-half topology, which provides additional structural information during training. For example, when the given graph is homophily-prone, the found topology provides additional information about the inter-class counterparts of target nodes. Then, with \ourmethod, target nodes can be avoided from being similar to these inter-class peers and thus improve their embedding quality. In contrast, the missing-half topology provides additional homophily-prone structural information for heterophily-prone datasets.


\subsection{Ablation Study}
We conduct an ablation study to evaluate the effectiveness of each component of \ourmethod.
Specifically, we consider seven variants including $\ourmethod_{\alpha = 0}$, $\ourmethod_{\beta = 0}$, $\ourmethod_{\gamma = 0}$, $\ourmethod_{\delta = 0}$, $\ourmethod_{\beta}$, $\ourmethod_{\gamma}$, and $\ourmethod_{\delta}$. The first four variants mean $\ourmethod$ without one component, and the latter four variants mean $\ourmethod$ with only one component, e.g., $\ourmethod_{\beta}$ means $\alpha, \gamma, \delta = 0$ and $\beta \neq 0$. The experiment result is presented in Table~\ref{tab:ablation}.

Interestingly, \hlt{$\ourmethod_{\beta}$ standalone already achieves SOTA results on heterophily-prone datasets (i.e., Chameleon, Squirrel and Actor) using only the learnt homophily-prone topology}. On the other hand, $\ourmethod_{\gamma}$ standalone achieves competitive results with less than 5\% gap apart from the best result on most homophily-prone datasets. In addition, $\ourmethod_{\delta = 0}$ reaches the best performance of $\ourmethod$ in only three out of eight datasets. This indicates the consideration of extended heterophily-prone neighbours is useful.

\subsection{Evaluation on the Found Missing Half Topology}
We evaluate the quality of the found missing-half topology based on the homophily ratio. Here we provide the definition of homophily ratio:
\begin{definition}
The homophily ratio of a graph, which evaluates the proportion of homophilic connections can be calculated as follows:
\begin{equation}
    \mathcal{H}(\mathcal{G}) = \frac{\sum_{(i,j) \in \mathcal{E}_{\mathcal{G}}} (Y_{v_i} == Y_{v_j})}{|\mathcal{E}_{\mathcal{G}}|},
\end{equation}
where 
$\mathcal{E}_{\mathcal{G}}$ is the set of edges in $\mathcal{G}$, $|\cdot|$ indicates the size of a set, $Y_{v_i}$ represents the label for node $i$, and $==$ evaluates whether two nodes have the identical label and assign 1 to intra-class edges, otherwise 0.
\label{def:homo ratio}
\end{definition}

\begin{figure}[t]
    \centering
\includegraphics[scale = 0.35]
{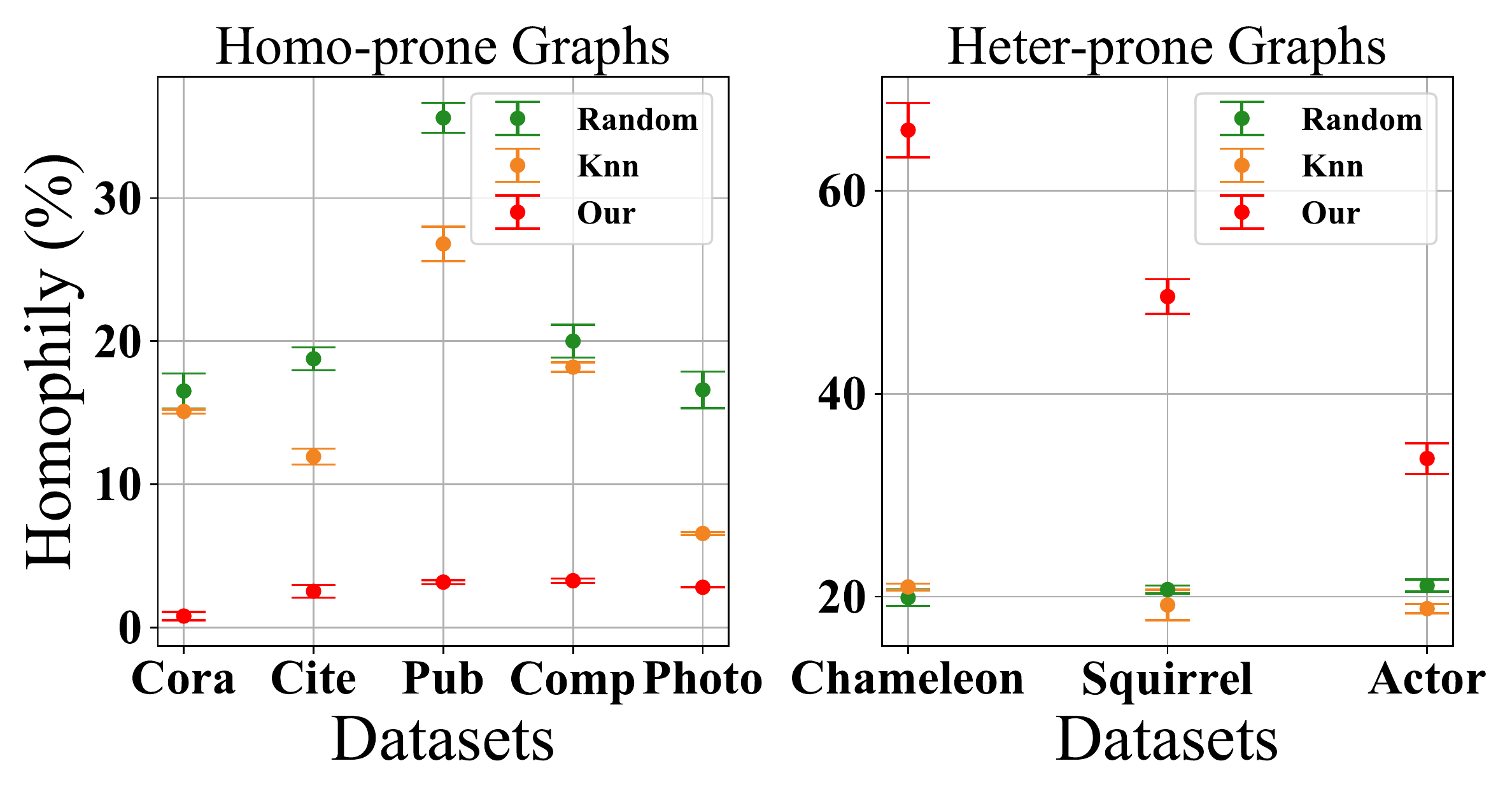}
\vspace{-5mm}
    \caption{Generated graphs homophily ratio comparison among our method in section~\ref{sec:method_gcl}, KNN and randomly generated graph. The two boundary line of data points in the figure indicates their standard deviation. `Cite', 'Pub' and `Comp' respectively mean CiteSeer, Pubmed, and Amazon Computers datasets.}
    \label{fig:comp_homophily}
\vspace{-3mm}
\end{figure}

In particular, a learnt heterophily-prone topology should have a low homophily ratio, while a learnt homophily-prone topology should have a high homophily ratio. This indicates the learnt topology is accurate.

We compare our graph complementation method in section~\ref{sec:method_gcl} with graphs generated with the KNN algorithm and randomly generated graphs using Erdős–Rényi model~\cite{erdHos1960evolution} in terms of homophily ratio. Given two parameters $N$ (i.e., number of nodes) and $E$ (i.e., number of edges), the Erdős–Rényi model uniformly samples $E$ edges from all possible node pairs in a graph to create random graphs. In our experiment, all generated graphs have the same number of edges. For our method and KNN, we generate complementary heterophily-prone topology for homophily-prone datasets and do the opposite for heterophily-prone datasets as introduced in section~\ref{par:graph compl}. For all methods, we generate the graphs ten times and calculate their homophily ratio. 
The mean and standard deviation of the homophily ratio on the generated graphs are presented in Figure~\ref{fig:comp_homophily}.

As shown in the figure, on homophily-prone datasets (i.e., Cora, Citeseer, Pubmed, Computers, and Photo), our method has a significantly lower homophily ratio compared with generated graphs from the other two methods. Graphs generated with our method for heterophily-prone graphs (i.e., Chameleon, Squirrel, and Actor), on the other hand, have significantly higher homophily ratios than the other two baselines. For example, the homophily ratio for the Chameleon graph is only 23\%. However, the learnt homophily-prone topology with our method for this dataset has its homophily ratio reaching 67\%.

\subsection{Scalability Extension}
\label{sec:scalability}

The scalability of \ourmethod\ can be enhanced by incroprating node sampling during the graph complementation step. We have successfully extend \ourmethod\ to a large-scale dataset Ogbn-arxiv, containing 169,343 nodes and 1,166,243 edges. The experiment results of \ourmethod\ 
and five baseline counterparts are shown in Table \ref{tab:large_exp}. Notably,  \ourmethod\ outperforms the five baselines, achieving the best results. 

\begin{table}[t]
\footnotesize
\centering
\setlength{\tabcolsep}{3.5pt} 
\caption{\ourmethod\ performance compared with 5 selected baselines on Ogbn-arxiv.}
\begin{tabular}{lccccc|c}
   \toprule
   Model      & GCN   & GAT   & APPNP & JKNET & SAGE & \textbf{GOAL} \\
   \midrule
   Ogbn-arxiv & 70.60 & 70.52 & 70.56 & 70.90 & 70.34     & 71.25 \\
   \bottomrule
   \end{tabular}
\label{tab:large_exp}
\vspace{-3mm}
\end{table}

\begin{table}[t]
\centering
\footnotesize
\setlength{\tabcolsep}{3.5pt} 
\caption{$\ourmethod_{at}$ performance compared with $\ourmethod$, the best and the second-best baseline performance on five selected datasets.}
\begin{tabular}{lccccc}
\toprule
Dataset & Pubmed & Computer & Chameleon & Squirrel & Actor \\ \hline
Best & 89.05 & 89.77 & 67.96 & 51.80 & 35.85 \\
Second & 88.64 & 88.54 & 64.63 & 49.52 & 31.76 \\ 
\ourmethod & 89.25 & 91.33 & 71.65 & 60.53 & 36.46 \\ \hline
\textbf{$\textbf{\ourmethod}_{at}$} & 88.37 & 89.85 & 69.58 & 61.76 & 35.52 \\
\bottomrule
\end{tabular}
\label{tab:GOAL_autotune}
\end{table}

During graph complementation, node pair comparisons occur when calculating both the grouping loss and learning-to-rank loss. To simplify the computation of grouping loss, we can sample a fixed number of intra-class and inter-class node pairs rather than computing the loss for all possible node pairs. For learning-to-rank loss, we first need to construct ranking lists for each node in the training set. Rather than finding similarities between all node pairs for ranking list construction, we can sample a node subset and calculate similarities solely between the training set nodes and the sampled nodes. By constructing the ranking list using the sampled node set, the learning-to-rank loss can be easily computed, thereby improving the scalability of GOAL effectively.

\subsection{Automatic Prameter-Tuning}
\label{sec:auto_tune}
Our proposed Complemented Graph Convolution method may be hard to tune, as it has four hyper-parameters: $\alpha$, $\beta$, $\gamma$, and $\delta$ to control the weight of each convolution component in Equation \ref{eq:agc conv}. To alleviate this problem, we also introduce an automatic parameter-tuning version of \ourmethod, namely, $\ourmethod_{at}$.  $\ourmethod_{at}$can automatically tune these four parameters by transforming these parameters into learnable parameters. Thus, as opposed to hand-tuned hyperparameters, they can be updated automatically during model training via gradient descent. We have conducted experiments on a selected set of 5 datasets, and the results are presented in Table \ref{tab:GOAL_autotune}.

The table demonstrates that $\ourmethod_{at}$\ significantly outperforms the second-best baseline on all datasets except for PubMed. Furthermore, $\ourmethod_{at}$\ achieves state-of-the-art performance on three out of the five datasets. Remarkably, $\ourmethod_{at}$\ even surpasses the original model on the Squirrel dataset by a margin of 1.2\%.


\section{Future Work}
\label{sec:future}

In this paper, we introduce Graph Complementary Learning to address the ``missing-half'' topology problem. Though we have achieved promising results, we can still consider the following aspects as future directions.
For example, could we improve the graph complementation by using a more advanced encoder instead of MLP or adding more losses? Could we design a better graph convolution with other objectives? More importantly, graph complementary learning can be employed in various real-world graphs, e.g., social networks, to enrich their structural information.

\section*{Acknowledgements}
\label{sec:ack}
This research was supported by an Australian Research Council (ARC) Future Fellowship
(FT210100097).

\bibliography{example_paper}
\bibliographystyle{icml2023}

\newpage
\appendix
\onecolumn


\section{Eigenvalues of ($\textbf{I} + \hat{\textbf{A}}_{t}\hat{\textbf{A}}_{o}$)}
\label{sec:appendix eigen}

\begin{table}[h]
\centering
\small
\caption{The range of eigenvalues for $\textbf{I} + \hat{\textbf{A}}_{t}\hat{\textbf{A}}_{o}$ on eight datasets}
\begin{tabular}{l|llllllll} 
\toprule
\textbf{Dataset} & \textbf{Cora} & \textbf{Citeseer} & \textbf{Pubmed} & \textbf{Computers} & \textbf{Photo}     & \textbf{Chameleon} & \textbf{Squirrel} & \textbf{Actor}  \\ 
\midrule
\textbf{Range}   & {[}0.2151, 2] & {[}0.2018, 2]     & {[}0.043, 2]    & {[}0.3193, 2]      & {[}0.2860, 1.9353] & {[}0.6252, 2]      & {[}0.7637, 2]     & {[}0.8435, 2]   \\
\bottomrule
\end{tabular}
\label{tab:eigen}
\end{table}
In addition to the theoretical evidence of $\textbf{I} + \hat{\textbf{A}}_{to}$ is positive semi-definite when solving $\mathcal{O}_{c}$, we also conduct empirical evaluation on the eigenvalues of $\textbf{I} + \hat{\textbf{A}}_{to}$.
Specifically, we calculate the eigenvalues range for $\textbf{I} + \hat{\textbf{A}}_{to}$ on eight datasets and the results are shown in Table ~\ref{tab:eigen}.
From the table, we can see all eigenvalues in these datasets are positive, and thus $\textbf{I} + \hat{\textbf{A}}_{to}$ are positive semi-definitive matrices in all datasets.

\section{Interpreting $\ourconvmethod$ from the Perspective of Graph Spectral Theory}
\label{sec:appendix spectral}
To illustrate \ourconvmethod\ from the perspective of graph spectral theory, we first introduce some background knowledge of bridging spectral GNNs and graph convolution forms, followed by transferring two spectral filters to convolutions in \ourconvmethod. After demonstrating the functionality of these graph spectral filters, we present the proof for Proposition~\ref{prop:agc conv spectral}.

\subsection{Bridging Spectral GNNs with Graph Convolution}
\label{sec: bridging}
Spectral GNNs use spectral graph theory as the foundation. In this framework, signals on graphs are filtered using the eigendecomposition of graph Laplacian $\hat{\textbf{L}}$. Since $\hat{\textbf{L}}$ is positive semidefinite, it can be decomposed to $\hat{\textbf{L}} = \textbf{U}^T\Lambda\textbf{U}$. Here $\Lambda = diag(\lambda)$ ($\lambda$ represents the vector of eigenvalues) and $\textbf{U}$ indicates the matrix of eigenvectors. Then, we define the graph Fourier transform of a signal on a graph as $\tilde{\textbf{X}} = \textbf{U}^T\textbf{X} \in \mathbf{R}^{N \times D}$, and the inverse transform is $\textbf{X} = \textbf{U}\tilde{\textbf{X}}$. By transposing the convolution theorem to graphs, the graph convolution kernel with spectral filtering can be defined as:
\begin{equation}
\label{eq:gconv kernel}
    \mathcal{A} = \textbf{U}diag(\textbf{F}(\lambda))\textbf{U}^T,
\end{equation}
where $\textbf{F}(\lambda)$ is the  filter function for filtering eigenvalues $\lambda$. Then, a uniform formula of graph convolution bridging the spectral-based approaches and spatial-based approaches can be defined below~\cite{balcilar2020bridging}:
\begin{equation}
\label{eq:message passing form}
    \textbf{H}^{l + 1} = \sigma(\sum^K_{k=1}\mathcal{A}_k\textbf{H}^{l}\textbf{W}^{(l,k)}),
\end{equation}
where $K$ is the number of filter functions, $\sigma(\cdot)$ is the non-linear activate function, $\textbf{W}^{(l,k)} \in \mathbb{R}^{D' \times D'}$ ($\textbf{W}^{(0,k)} \in \mathbb{R}^{D \times D'})$ is a learnable weight matrix at layer $l$ with $k$-th graph convolution kernel, $\textbf{H}^{l} \in \mathbb{R}^{N\times D'}$ is node representations at $l$-th layer ($\textbf{H}^{(0)} = \textbf{X}$). With Equation \ref{eq:gconv kernel} and \ref{eq:message passing form}, we can design various spectral-based methods using different filter functions and obtain their graph convolution. Taking ChebyNet~\cite{defferrard2016convolutional} as an example, it uses Chebyshev polynomial to form an orthogonal basis and takes them as a series of filter functions:  $\textbf{F}_1(\lambda) = 1, \textbf{F}_2(\lambda) = \frac{2\lambda}{\lambda_{max}}  - 1, \cdots$, $\textbf{F}_k(\lambda) = 2\textbf{F}_2(\lambda)\textbf{F}_{k-1}(\lambda) - \textbf{F}_{k-2}(\lambda)$. In this case, $\mathcal{A}_k$ in Equation \ref{eq:message passing form} corresponds to the convolution kernel of each $\textbf{F}_k(\lambda)$. Another example is GCN~\cite{kipf2016semi}, which can be considered as the simplification of ChebNet with first-order approximation with $K =2$ and $\lambda_{max} = 2$. The filter function of GCN can be approximated by $\textbf{F}(\lambda) = 1 - \lambda \frac{\overline{d}}{(1 + \overline{d})}$, where $\overline{d}$ is the average degree of nodes~\cite{wu2022beyond}. Thus, we can observe that GCN only consists of a low-pass filter.

\subsection{Interpreting \ourconvmethod\ with Spectral Filters}
\label{sec:appendix interpret}

This section first demonstrates how specific spectral filters are consistent with operations in \ourconvmethod\, followed by pointing out the desired graph topology of specific spectral filters.




\subsubsection{Low-pass Filter}

\textbf{Effect.} Existing GNN approaches primarily use low-pass filters, e.g., GCN, SGC, and APPNP~\cite{he2021bernnet}. Here, we analyse a typical linear low-pass filter:
\begin{equation}
    \textbf{F}_{low}(\lambda) = 1 - \lambda,
    \label{eq:filter low}
\end{equation}
Based on Equation~\ref{eq:gconv kernel} and Equation~\ref{eq:message passing form}, we can infer its graph convolution form is:
\begin{align}
    \begin{aligned}
            \mathcal{A} &=  \hat{\textbf{A}},
        \\ \textbf{H}^{l + 1}  &= \sigma(\hat{\textbf{A}}\textbf{H}^{l}\textbf{W}^{l}).
\end{aligned}
\label{eq:gconv low}
\end{align}



\textbf{Desired Topology.}
From Equation~\ref{eq:gconv low}, we can see it is consistent with the graph convolution (i.e., Equation~\ref{eq:homo conv}) derived from objective $\mathcal{O}_o$ in section~\ref{sec:optimisation agc}. Accordingly, we can see that the linear low-pass filter minimises the Euclidean distance between two connected nodes. This indicates the filtering process relies heavily on the homophily-prone edges of a given graph. When homophily-prone edges are dominant in a graph, same-class nodes can be clustered using low-pass filters. In contrast, if a graph is highly heterophily-prone, these low-pass filters tend to maximise the similarity between inter-class nodes and lead to poor performance. This explains why low-pass-based GNNs such as GCN~\cite{kipf2016semi} and GAT~\cite{velivckovic2017graph} work poorly on heterophily-prone graphs. Thus, the desired topology for low-pass filters is homophily-prone graphs.

\subsubsection{High-pass Filter}
\textbf{Effect.} Some existing works incorporate linear high-pass filters to capture information for the high-frequency band of graph signals \cite{fagcn2021,wu2022beyond}. Here, we consider the following linear high-pass filter:
\begin{equation}
    \textbf{F}_{high}(\lambda) = \lambda -1,
\label{eq:filter high}
\end{equation}
According to Equations~\ref{eq:gconv kernel} and \ref{eq:message passing form}, the graph convolution form of Equation~\ref{eq:filter high} can be defined as:
\begin{align}
    \begin{aligned}
        \mathcal{A} &=  - \hat{\textbf{A}}, \\
        \textbf{H}^{l + 1} &= \sigma(-\hat{\textbf{A}}\textbf{H}^l\textbf{W}^l),
    \end{aligned}
        \label{eq:gconv high}
\end{align}

\textbf{Desired Topology.} Equation~\ref{eq:gconv high} is identical to the graph convolution derived from $\mathcal{O}_t$ (i.e.,Equation~\ref{eq:heter conv}).
$\mathcal{A} =  - \hat{\textbf{A}}$ indicates that the desired topology for linear high-pass filters should be heterophily-prone graphs because $\mathcal{A}$ can maximise the dissimilarity between two connected nodes. 
If the input topology is homophily-prone, two intra-class nodes would be separated, which is undesirable. Conversely, a heterophily-prone input topology can improve target node embeddings by setting them apart from inter-class nodes. Thus, the desired input topology for linear high-pass filters should be heterophily-prone graphs.

\subsection{Overall Analysis}
\label{sec:appendix overall analysis}
To validate that the induced desired topology is consistent with corresponding filters, we conduct experiments using synthetic datasets, and the result is presented in the figure below:
\begin{figure}[h]
    \centering
\includegraphics[scale = 0.4]{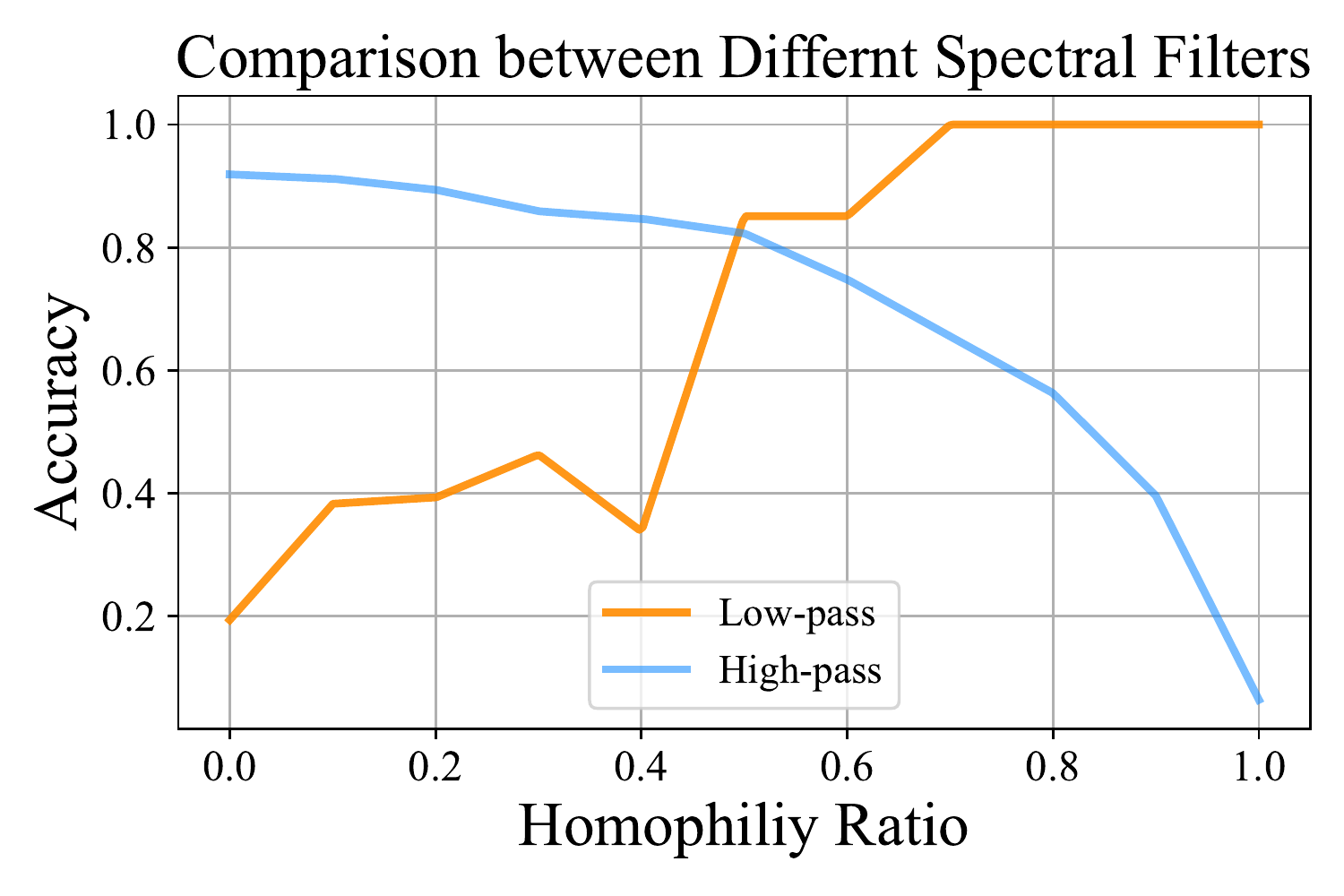}
    \caption{Comparison between Low- and High-pass Spectral Filters on a synthetic dataset.}
    \label{fig:compare different filters}
\end{figure}

The synthetic datasets are built based on the Cora dataset. Specifically, we delete all the edges in the Cora dataset and add synthetic connections to it. These connections have a homophily ratio ranging from 0 to 1. The definition of homophily ratio is provided in Definition~\ref{def:homo ratio}. From Figure~\ref{fig:compare different filters}, we can observe that the experiment result is consistent with our findings. 

When increasing the homophily ratio, the performance of the high-pass filter is on the decline, whereas the low-pass filter is on the rise. Accordingly, a high-pass filter performs well when a graph has a low homophily ratio. A Low-pass filter, on the other hand, prefers graphs with a high homophily ratio.
Thus, it is reasonable to apply the low-pass filter to $\hat{\textbf{A}}_o$ and the high-pass filter to $\hat{\textbf{A}}_t$ and $\hat{\textbf{A}}_{to}$. This is because $\hat{\textbf{A}}_o$ is homophily-prone, while the other two are heterophily-prone.

\subsection{Proof of Proposition~\ref{prop:agc conv spectral}}
\label{sec:appendix proof}
\begin{proof}
  \ourconvmethod\ can be considered as the combination of a linear low-pass filter and two high-pass filters.
  (1) As analysed in section~\ref{sec:appendix interpret}, the graph convolution derived from optimising $\mathcal{O}_o$ is identical to that from the defined low-pass filter $\textbf{F}_{low}(\lambda) = 1 - \lambda$ with respect to $\hat{\textbf{A}}_o$. 
  (2) In addition, the graph convolution obtained from the defined high-pass filter $\textbf{F}_{high}(\lambda) = \lambda -1$ concerning $\hat{\textbf{A}}_t$ is the same with that from optimising $\mathcal{O}_t$. 
  Similarly, the same high-pass filter concerning $\hat{\textbf{A}}_{to}$ and optimising $\mathcal{O}_c$ have the same graph convolution form.
  Thus, combining $\mathcal{O}_o$, $\mathcal{O}_t$ and $\mathcal{O}_c$ together is equivalent to applying the low-pass filter $\textbf{F}_{low}(\lambda) = 1 - \lambda$ to the Laplacian of $\hat{\textbf{A}}_o$ and applying the high-pass filter $\textbf{F}_{high}(\lambda) = \lambda -1$ to the Laplacian of $\hat{\textbf{A}}_o$ and $\hat{\textbf{A}}_{to}$, respectively.
\end{proof}

\section{Dataset Statistics}
\label{sec:appendix exp}
This section presents the dataset statistics. Specifically, we adopt eight datasets in the experiment, which are Cora, \hlt{Citeseer, Pubmed}, Computers, Photo, Chameleon, and Squirrel. Their number of node $N$, number of edges $E$, number of classes $C$, number of feature dimension $D$ are summarised in the table below:
\begin{table*}[h]
\caption{Dataset statistics}
\label{tab:dataset}
\small
\centering
\begin{tabular}{l|llllllll} 
\toprule
                  & \textbf{Cora}  & \textbf{Citeseer} & \textbf{Pubmed}& \textbf{Computers} & \textbf{Photo}   & \textbf{Chameleon} & \textbf{Squirrel} &  \textbf{Actor} \\ 
\hline
\textbf{Nodes $N$}       & 2,708 & 3,327    & 19,717 & 13,752    & 7,650   & 2,277     & 5,201 & 7,600    \\
\textbf{Edges $E$}       & 5,278 & 4,552    & 44,324 & 245,861   & 119,081 & 31,371    & 198,353 & 26,659  \\
\textbf{Classes $C$ }    & 7     & 6        & 5      & 10        & 8       & 5         & 5  &  5        \\
\textbf{Feature $D$} & 1,433 & 3,703    & 500    & 767       & 745     & 2,325     & 2,089  & 932   \\
\bottomrule
\end{tabular}
\end{table*}

\end{document}